\documentclass[lettersize,journal]{IEEEtran}

\usepackage{mathrsfs}
\usepackage{amsmath}
\usepackage{amsfonts}
\usepackage{amsthm}
\usepackage{amssymb}
\usepackage{graphicx}
\usepackage{subfigure}
\usepackage{indentfirst}
\usepackage{array}
\usepackage{cite}
\usepackage{enumerate}
\usepackage{bm}
\usepackage[linesnumbered,ruled,vlined]{algorithm2e}
\usepackage{algorithmic}
\usepackage{multirow}
\usepackage{epstopdf}
\usepackage{verbatim}
\usepackage{stfloats}
\usepackage{color}
\usepackage{caption}
\usepackage{booktabs}
\usepackage{footnote}

\SetKwComment{Comment}{//}{}

\begin{document}

\title{Iterative Joint Detection of Kalman Filter and Channel Decoder for Sensor-to-Controller Link in Wireless Networked Control Systems}

\author{Jinnan Piao, Dong Li, Yiming Sun, Zhibo Li, Ming Yang, and Xueting Yu

\vspace{-0em}

\thanks{
This work is supported in part by the National Key R\&D Program of China under Grant 2024YFF0509700, in part by the National Natural Science Foundation of China under Grant 62201562, 62303449, and 92367301, in part by the Liaoning Provincial Natural Science Foundation of China under Grant 2024--BSBA--51, and in part by the Fundamental Research Project of SIA under Grant 2023JC1K09. (\emph{Corresponding Author: Dong Li})
}
\thanks{
The authors are with the State Key Laboratory of Robotics, Shenyang Institute of Automation, Chinese Academy of Sciences, Shenyang 110016, China.
(e-mail: piaojinnan@sia.cn; lidong@sia.cn; sunyiming@sia.cn; lizhibo@sia.cn; yangming@sia.cn; yuxueting@sia.cn)}
}

\maketitle

\begin{abstract}

In this letter, we propose an iterative joint detection algorithm of Kalman filter (KF) and channel decoder for the sensor-to-controller link of wireless networked control systems, which utilizes the prior information of control systems to improve control and communication performance. In this algorithm, we first use the KF to estimate the probability density of the control system outputs and calculate the prior probability of received signals to assist the decoder. Then, we traverse the possible outputs of the control system to update the prior probability to implement iterative detection. The simulation results show that the prior information and the iterative structure can reduce the block error rate performance of communications while improving the root mean square error performance of controls.

\end{abstract}


\section{Introduction}
\IEEEPARstart{W}{ireless} networked control systems (WNCSs) \cite{NCS_Survey, WNCS_Survey, 9547791} are control systems  with the components, i.e., controllers, sensors and actuators, distributed and connected via wireless communication channels, where transmission errors tightly relate to the control stability and performance.
Kalman filter (KF) plays a fundamental role in estimating system states when transmission errors occur \cite{KalmanLQR}. The major researches of KF model the wireless channels as independent and identically distributed (i.i.d.) Bernoulli processes \cite{KalmanBernoulliOriginal, KalmanBernoulli1, KalmanBernoulli2} or Markov processes \cite{KalmanMarkov1, KalmanMarkov2, KalmanMarkov3} to reduce the influence of transmission errors on control performance.

From the view of communications, channel codes can effectively reduce transmission errors \cite{LinShuBook} and a simple on-off error control coding scheme can improve the control quality \cite{WNCSchannelCodingOnOff}. To ensure a wide range of application scenarios, channel codes are generally designed assuming that the transmitted bits follow a uniform distribution and no prior information is considered.
To improve the control and communication performance with the prior information of control systems, a maximum a posteriori (MAP) receiver for each element of system states with cyclic redundancy check (CRC) codes is proposed in \cite{WNCSchannelCodingCRC}, which exhibits potential in optimizing the block error rate (BLER) performance of communications and the root mean square error (RMSE) performance of controls with the prior information.
However, the prior information directly calculated by the system states cannot be used in channel decoder.

To utilize the prior information for channel decoder, we propose an iterative joint detection algorithm in this letter, which exchanges the prior information of quantized bits and the decoded probabilities of outputs between KF and channel decoder.
In the algorithm, we first use the KF to estimate the probability density of the predicted system states.
Then, the probability density is transformed into the prior logarithmic likelihood ratios (LLRs) of quantized bits to assist the decoder and obtain the decoded probabilities of outputs.
Finally, the possible outputs are traversed to update the prior LLRs of quantized bits to implement iterative detection.
The simulation results show that the prior information and the iterative structure can reduce the BLER performance of communications while improving the RMSE performance of controls, which shows the advantage of the joint design of controls and communications.

\emph{Notation Conventions}: In this letter, the lowercase letters, e.g., $x$, are used to denote scalars. The bold lowercase letters, e.g., ${\mathbf{x}}$, are used to denote vectors.
Notation $x_i$ denotes the $i$-th element of ${\mathbf{x}}$.
The sets are denoted by calligraphic characters, e.g., $\mathcal{X}$, and the notation $|\mathcal{X}|$ denotes the cardinality of $\mathcal{X}$.
The bold capital letters, e.g., $\mathbf{X}$, are used to denote matrices.
Throughout this paper, $\mathbf 0$ denotes an all-zero vector.
${\mathbb{R}}$ represents the real number field.
$\left[\!\left[ N
 \right]\!\right]$ denotes the set $\left\{1,2,\cdots,N\right\}$.

\section{System Model}

\begin{figure*}[t]
\setlength{\abovecaptionskip}{0.cm}
\setlength{\belowcaptionskip}{-0.cm}
  \centering{\includegraphics[scale=0.7]{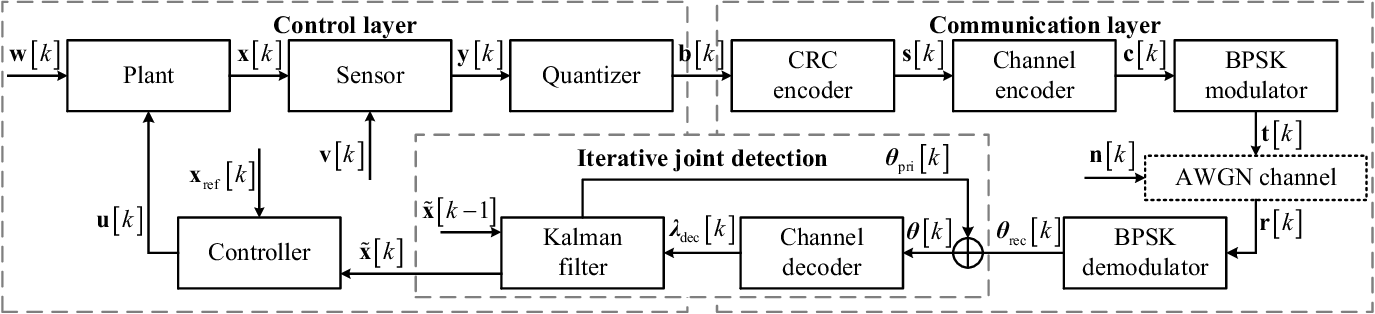}}
  \caption{The system structure of WNCSs with control layer, communication layer and iterative joint detection.}\label{FigSystemModel}
  \vspace{-1em}
\end{figure*}

The considered system structure of WNCSs in this letter is shown in Fig. \ref{FigSystemModel}, which consists of control layer and communication layer.

In the control layer, we consider a discrete linear time-invariant system as
\begin{align}
  \mathbf{x}\left[ k+1 \right]&=\mathbf{Ax}\left[ k \right]+\mathbf{Bu}\left[ k \right]+\mathbf{w}\left[ k \right],  \label{EqPlant} \\
 \mathbf{y}\left[ k \right]&=\mathbf{Cx}\left[ k \right]+\mathbf{v}\left[ k \right], \label{EqSensor}
\end{align}
where $\mathbf{x}\left[ k \right] \in {{\mathbb{R}}^{{{N}_{x}}}}$, $\mathbf{u}\left[ k \right] \in {{\mathbb{R}}^{{{N}_{u}}}}$ and $\mathbf{y}\left[ k \right]\in {{\mathbb{R}}^{{{N}_{y}}}}$ are the state vector, the input vector and the output vector at time index $k$, respectively. $\mathbf{A}\in {{\mathbb{R}}^{{{N}_{x}}\times {{N}_{x}}}}$, $\mathbf{B}\in {{\mathbb{R}}^{{{N}_{x}}\times {{N}_{u}}}}$ and $\mathbf{C}\in {{\mathbb{R}}^{{{N}_{y}}\times {{N}_{x}}}}$ are the known system parameter matrices. $\mathbf{w}\left[ k \right]\in {{\mathbb{R}}^{{{N}_{x}}}}$ and $\mathbf{v}\left[ k \right]\in {{\mathbb{R}}^{{{N}_{y}}}}$ are Gaussian noises with zero means and covariance matrices ${\mathbf W}$ and ${\mathbf V}$, respectively.

Since our work focuses on the joint detection of communications and controls rather than the quantization design, we select the simple and widely adopted uniform quantization for the sensor outputs. Each element of $\mathbf{y}\left[ k \right]$ is quantized by an $n$-bit uniform quantizer with the quantization range $\left[ -Z, Z \right)$.
Defining $\alpha \left( x \right)\triangleq {\hat z}_l = 0.5\left( {{z}_{l}}+{{z}_{l+1}} \right)$ as the midpoint of quantized interval $x\in \left[ {{z}_{l}},{{z}_{l+1}} \right)$,
$\beta \left( x \right)\triangleq l$ as the index of quantized interval with $z_0 = -Z$, $z_l = z_{l-1} + \Delta$ and $\Delta ={Z}/{\left( {2}^{n-1} \right)}$, and ${\mathbf b}^m = \left[b_{n-1}^m, \cdots, b_1^m, b_0^m\right]$ as the binary representation of $m$ with the most and the least significant bits $b_{n-1}^m$ and $b_0^m$,
the $N_y$-length quantized vector and the $\left(N_yn\right)$-bit vector are
$\mathbf{q}\left[ k \right]=\left[ \alpha \left( {{y}_{1}}\left[ k \right] \right),\alpha \left( {{y}_{2}}\left[ k \right] \right),\cdots ,\alpha \left( {{y}_{{{N}_{y}}}}\left[ k \right] \right) \right]$ and
$\mathbf{b}\left[ k \right]=\left[ {{\mathbf{b}}^{\beta \left( {{y}_{1}}\left[ k \right] \right)}},{{\mathbf{b}}^{\beta \left( {{y}_{2}}\left[ k \right] \right)}},\cdots ,{{\mathbf{b}}^{\beta \left( {{y}_{{{N}_{y}}}}\left[ k \right] \right)}} \right]$, respectively.
$\mathbf{b}\left[ k \right]$ is sent to the communication layer.

After the iterative joint detection, the estimation $\mathbf{\hat{q}}\left[ k \right]$ of $\mathbf{{q}}\left[ k \right]$ is obtained and
KF is used to estimate the system states $\mathbf{\tilde{x}}\left[ k \right]$ and the covariance matrix $\mathbf{\tilde{P}}\left[ k \right]$ in Algorithm \ref{AlgorithmKF} with the inputs $\mathbf{\tilde{x}}\left[ k-1 \right]$, $\mathbf{\tilde{P}}\left[ k-1 \right]$, $\mathbf{\hat{q}}\left[ k \right]$ and $f_{\text{crc}}$, where $f_{\text{crc}} = 0$ if the decoded bits pass CRC, and $f_{\text{crc}} = 1$ otherwise.
Then, controller uses $\mathbf{\tilde{x}}\left[ k \right]$ and the reference state vector ${{\mathbf{x}}_{\text{ref}}}\left[ k \right] \in {\mathbb R}^{N_x}$ to calculate the input vector as
\begin{equation}\label{EqInputVector}
\mathbf{u}\left[ k \right] = \mathbf{K}_\text {con}\left( {{\mathbf{x}}_{\text{ref}}}\left[ k \right]- \mathbf{\tilde{x}}\left[ k \right]\right),
\end{equation}
where $\mathbf{K}_\text{con} \in {\mathbb R}^{N_u \times N_x}$ is the controller gain matrix.

\begin{algorithm}[t]
\setlength{\abovecaptionskip}{0.cm}
\setlength{\belowcaptionskip}{-0.cm}
\caption{Kalman Filter: {($\mathbf{\tilde{x}}\left[ k \right]$, $\mathbf{\tilde{P}}\left[ k \right]$) = {\sf KF}($\mathbf{\tilde{x}}\left[ k-1 \right]$, $\mathbf{\tilde{P}}\left[ k-1 \right]$, $\mathbf{\hat{q}}\left[ k \right]$, $f_{\text{crc}}$)}}\label{AlgorithmKF}

  $\mathbf{\tilde{x}}_-\left[ k \right] =  \mathbf{A}\mathbf{\tilde{x}}\left[ k-1 \right] + \mathbf{Bu}\left[ k - 1 \right]$ \;

 $\mathbf{\tilde{P}}_-\left[ k \right]  =\mathbf{A\tilde{P}}\left[ k-1 \right]{{\mathbf{A}}^{T}}+\mathbf{W}$ \;

 $\mathbf{K}\left[ k \right]=\mathbf{\tilde{P}}_-\left[ k \right]{{\mathbf{C}}^{T}}{{\left( \mathbf{C}\mathbf{\tilde{P}}_-\left[ k \right]{{\mathbf{C}}^{T}}+\mathbf{V} \right)}^{-1}}$ \;

 \If{$f_{\text{crc}}$ is $0$}
 {
    $\mathbf{\tilde{x}}\left[ k \right] = \mathbf{\tilde{x}}_-\left[ k \right]+\mathbf{K}\left[ k \right]\left( \mathbf{\hat{q}}\left[ k \right]-\mathbf{C}\mathbf{\tilde{x}}_-\left[ k \right] \right)$\;

    $\mathbf{\tilde{P}}\left[ k \right] = \left( \mathbf{I}-\mathbf{K}\left[ k \right]\mathbf{C} \right)\mathbf{\tilde{P}}_-\left[ k \right]$\;
 }
 \Else
 {
    $\mathbf{\tilde{x}}\left[ k \right] = \mathbf{\tilde{x}}_-\left[ k \right]$ and
    $\mathbf{\tilde{P}}\left[ k \right] = \mathbf{\tilde{P}}_-\left[ k \right]$\;
 }

\end{algorithm}

In the communication layer, the outer code is an $\left(N_{\text{out}}, K=N_yn\right)$ CRC code, the inner code is an $\left(N_{\text{in}}, N_{\text{out}}\right)$ channel code and the code rate is $R = K/N_{\text{in}}$.
$\mathbf{b}\left[k\right]$ is encoded into the CRC code by $\mathbf{s}\left[k\right] = \mathbf{b}\left[k\right]\mathbf{G}_{\text{out}}$ and $\mathbf{s}\left[k\right]$ is encoded into the
codeword by $\mathbf{c}\left[k\right] = \mathbf{s}\left[k\right]\mathbf{G}_{\text{in}}$, where $\mathbf{G}_{\text{out}}$ and $\mathbf{G}_{\text{in}}$ are the generator matrices of the CRC code and the channel code, respectively.
Each coded bit $c_i\left[k\right]$ is modulated into the transmitted signal by binary phase shift keying (BPSK), i.e., $t_i\left[k\right] = 1 - 2c_i\left[k\right]$.
The received vector is $\mathbf{r}\left[k\right] = \mathbf{t}\left[k\right] + \mathbf{n}\left[k\right]$,
where $n_i\left[k\right]$ is i.i.d. additive white Gaussian noise (AWGN) with zero mean and variance $\sigma^2$.
After BPSK demodulator, $\mathbf{r}\left[k\right]$ is transformed into the received LLR vector ${\bm \theta}_{\text{rec}}\left[k\right]$ with
\begin{equation}\label{EqLLRrec}
\theta_{\text{rec},i}\left[k\right] = \ln \frac{p \left( \left. {r_i}\left[k\right] \right|{{c}_{i}}\left[k\right]=0 \right)}{p \left( \left. {{r}_{i}}\left[k\right] \right|{{c}_{i}}\left[k\right]=1 \right)}= \frac{2{{r}_{i}}\left[k\right]}{{{\sigma }^{2}}},i\in\left[\!\left[ N_{\text{in}}
 \right]\!\right].
\end{equation}
Then, ${\bm \theta}_{\text{rec}}\left[k\right]$ is sent to the iterative joint detection.

\section{Iterative Joint Detection}

\begin{figure}[t]
\setlength{\abovecaptionskip}{0.cm}
\setlength{\belowcaptionskip}{-0.cm}
  \centering{\includegraphics[scale=0.65]{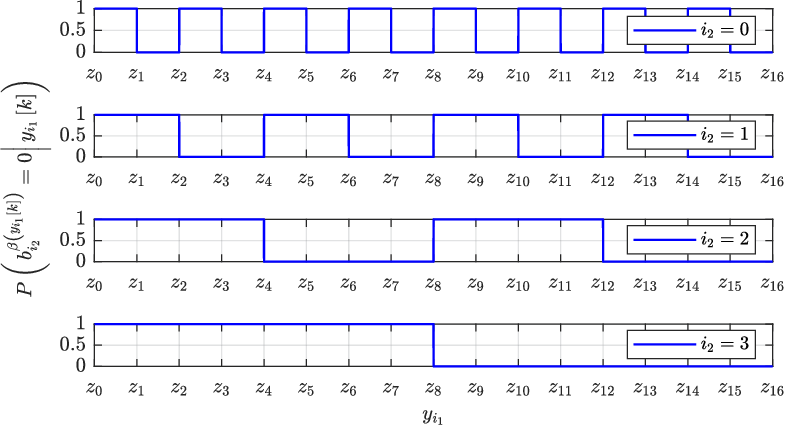}}
  \caption{An example of $P\left( \left. b_{i_2}^{\beta \left( y_{i_1}\left[k\right] \right)} =0\right|{{y}_{i_1}\left[k\right]} \right)$ with $n = 4$.}\label{FigExample}
  \vspace{-1em}
\end{figure}

The structure of iterative joint detection is also provided in Fig. \ref{FigSystemModel}, which consists of KF and channel decoder.
The prior LLR vector ${\bm \theta}_{\text{pri}}\left[k\right]$ is sent from the KF to the channel decoder and the decoded LLR vector ${\bm \lambda}_{\text{dec}}\left[k\right]$ is sent oppositely, where
\begin{align}
{\theta}_{\text{pri},i}\left[k\right] & = \ln \frac{P \left( {{c}_{i}}\left[k\right]=0 \right)}{P \left( {{c}_{i}}\left[k\right]=1 \right)}, i\in\left[\!\left[ N_{\text{in}}
 \right]\!\right], \label{EqLLRpri}\\
{\lambda}_{\text{dec},j}\left[k\right] & = \ln \frac{P \left( \left. {{b}_{j}}\left[k\right]=0 \right|\mathbf{r}\left[k\right] \right)}{P \left( \left. {{b}_{j}}\left[k\right]=1 \right|\mathbf{r}\left[k\right] \right)}, j\in\left[\!\left[ K
 \right]\!\right]. \label{EqLLRdec}
\end{align}
The channel decoding algorithm, such as the belief propagation (BP) decoding of LDPC code \cite{LinShuBook}, is abstracted as a function $\left({\bm \lambda}_{\text{dec}}\left[k\right],f_{\text{crc}}\right)={\sf ChannelDecoding}\left({\bm \theta}\left[k\right]\right)$, where
\begin{equation}\label{EqLLR}
{{\theta }_{i}}\left[k\right] = \ln \frac{P \left( \left. {{c}_{i}}\left[k\right]=0 \right|{{r}_{i}}\left[k\right] \right)}{P \left( \left. {{c}_{i}}\left[k\right]=1 \right|{{r}_{i}}\left[k\right] \right)}={{\theta }_{\text{pri},i}}\left[k\right]+{{\theta }_{\text{rec},i}}\left[k\right].
\end{equation}
The proposed detection algorithm iteratively updates ${{\theta }_{\text{pri},i}}\left[k\right]$ by utilizing the prior information of $\mathbf{\tilde{x}}\left[ k-1 \right]$ to improve the control and communication performance, which is divided into the three steps as follows.

{\bf Step 1: Initializing} ${\bm \theta}_{\text{pri}}\left[k\right]$. We first use the KF to predict the system states $\mathbf{\hat x}\left[k\right]$ and the covariance matrix $\mathbf{\hat P}_x\left[k\right]$ as
\begin{equation}\label{EqIJDPredictXP}
\left(\mathbf{\hat x}\left[k\right], \mathbf{\hat P}_x\left[k\right]\right) = {\sf KF}\left(\mathbf{\tilde{x}}\left[ k-1 \right], \mathbf{\tilde{P}}\left[ k-1 \right], \mathbf{0}, 1\right).
\end{equation}
Then, the predicted sensor output is $
\mathbf{\hat y}\left[k\right] = \mathbf{C}\mathbf{\hat x}\left[k\right] \label{EqIJDPredictY}$ and the corresponding covariance matrix is $\mathbf{\hat P}_y\left[k\right] = \mathbf{C}\mathbf{\hat P}_x\left[k\right]\mathbf{C}^T + \mathbf{V}$.
Since $\mathbf{x}\left[k\right]$  follows a Gaussian distribution by \eqref{EqPlant}, $\mathbf{y}\left[k\right]$ follows the Gaussian distribution with mean $\mathbf{\hat y}\left[k\right]$ and covariance matrix $\mathbf{\hat P}_y\left[k\right]$ as
\begin{align}\label{EqProYgivenXk1}
\mathbf{y}\left[k\right] \sim \mathcal{N}\left(\mathbf{\hat y}\left[k\right], \mathbf{\hat P}_y\left[k\right]\right).
\end{align}

As ${\mathbf y}\left[k\right]$ is transformed into ${\mathbf q}\left[k\right]$ and ${\mathbf b}\left[k\right]$ by quantizer, the probability of quantized bit $b_{i_2}^{\beta \left( y_{i_1}\left[k\right] \right)} \in \left\{0, 1\right\}$, ${i_1} \in \left[\!\left[ N_y
 \right]\!\right]$, ${i_2} = n-1,\cdots,1,0$, is
\begin{equation}\label{EqProInfoBit}
\begin{gathered}
  P \left( {{b_{i_2}^{\beta \left( y_{i_1}\left[k\right] \right)}} } \right)
   = \int\limits_{y_{i_1}\left[k\right] \in \mathbb{R}} {p\left( {{b_{i_2}^{\beta \left( y_{i_1}\left[k\right] \right)},y_{i_1}\left[k\right]} } \right)} dy_{i_1}\left[k\right] \hfill \\
   = \int\limits_{y_{i_1}\left[k\right] \in \mathbb{R}} {P\left( {\left. b_{i_2}^{\beta \left( y_{i_1}\left[k\right] \right)} \right|{y_{i_1}\left[k\right]}} \right)p\left( {y_{i_1}\left[k\right] } \right)} dy_{i_1}\left[k\right], \hfill \\
\end{gathered}
\end{equation}
where
\begin{align}
&y_{i_1} \left[k\right] \sim {\mathcal N}\left({\hat y}_{i_1}\left[k\right], \sigma _{{\hat y}_{i_1}\left[k\right]}\right), \label{EqpYiXk1}\\
&\begin{aligned}
  & P\left( \left. b_{i_2}^{\beta \left( y_{i_1}\left[k\right] \right)}=0 \right|{{y}_{i_1}\left[k\right]} \right) \\
 & =\left\{ \begin{aligned}
  & 1,\text{if }{{y}_{i_1}\left[k\right]}\in \left[ {{z}_{l\times{{2}^{(i_2+1)}}}},{{z}_{l\times{{2}^{(i_2+1)}}+2^{i_2}\Delta}}  \right), \\
 & 0,\text{otherwise}, \\
\end{aligned} \right. \\
\end{aligned} \label{EqpbjYi0}
\end{align}
where $l=0,1,\cdots ,{{2}^{n-{i_2}-1}}-1$ and $\sigma _{{\hat y}_{i_1}\left[k\right]}^2$ is the ${i_1}$-th diagonal element of $\mathbf{\hat P}_y\left[k\right]$.
\eqref{EqpYiXk1} can be easily obtained by \eqref{EqProYgivenXk1}.
For \eqref{EqpbjYi0},
since $b_{i_2}^{\beta \left( y_{i_1}\left[k\right] \right)}$ is decided by quantizing $y_{i_1}\left[k\right]$ directly, we have $P\left( \left. b_{i_2}^{\beta \left( y_{i_1}\left[k\right] \right)} \right|{{y}_{i_1}\left[k\right]} \right) \in \left\{0,1\right\}$.
An example of $P\left( \left. b_{i_2}^{\beta \left( y_{i_1}\left[k\right] \right)} =0\right|{{y}_{i_1}\left[k\right]} \right)$ with $n = 4$ is provided in Fig. \ref{FigExample}.
In Fig. \ref{FigExample}, $P\left( \left. b_{i_2}^{\beta \left( y_{i_1}\left[k\right] \right)} =0\right|{{y}_{i_1}\left[k\right]} \right)$, $i_2 = 3,2,1,0$, changes from $0$ to $1$ or from $1$ to $0$ every ${{2}^{{{i}_{2}}}}$ quantized interval.

Then, we use $b_j\left[k\right]$ to represent the $j$-th element in ${\mathbf b}\left[k\right]$. There is a mapping between $b_j\left[k\right]$ and $b_{i_2}^{\beta \left( {{y}_{i_1}\left[k\right]} \right)}$, i.e.,
\begin{equation}\label{EqMapping}
j = {\mathcal M}\left(i_1, i_2\right) = (i_1 - 1)n+n-i_2=i_1n-i_2.
\end{equation}
Thus, $P \left( {{b_j}\left[k\right] } \right)$ can be calculated by \eqref{EqProInfoBit} and the prior LLR of $b_j\left[k\right]$ is
\begin{equation}\label{EqPriorLLRbjInitial}
{{\lambda }_{\text{pri},j}}\left[k\right] = \ln \left( {P\left( b_{{{i}_{2}}}^{\beta \left( {{y}_{{{i}_{1}}}}\left[ k \right] \right)}=0 \right)}/{P\left( b_{{{i}_{2}}}^{\beta \left( {{y}_{{{i}_{1}}}}\left[ k \right] \right)}=1 \right)}\; \right).
\end{equation}
Given ${\mathbf G} = \mathbf{G}_{\text{out}}\mathbf{G}_{\text{in}}$, the $j$-th row and the $i$-th column element $g_{j,i}$ of $\mathbf G$ and ${\mathcal{G}}\left( i \right)=\left\{ \left. j \right|{{g}_{j,i}}=1,j\in\left[\!\left[ K
 \right]\!\right] \right\},i\in \left[\!\left[ {{N}_{\text{in}}}
 \right]\!\right]$, we have
$
{{c}_{i}}\left[k\right]=\sum\nolimits_{j\in {\mathcal{G}}\left( i \right)}{{{b}_{j}\left[k\right]}}
$ and the prior LLR of $c_i\left[k\right]$ is
\begin{equation}\label{EqPriorLLRciCal}
{{\theta }_{\text{pri},i}}\left[k\right]=2{{\tanh }^{-1}}\left( \prod\limits_{j\in \mathsf{\mathcal{G}}\left( i \right)}{\tanh \left( \frac{1}{2}{{\lambda }_{\text{pri},j}\left[k\right]} \right)} \right).
\end{equation}

{\bf Step 2: Decoding ${\bm \theta}\left[k\right]$.} ${\theta }_{i}\left[k\right]$ is calculated by \eqref{EqLLR}, where $\theta_{\text{rec},i}\left[k\right]$ is decided by \eqref{EqLLRrec} and ${\theta }_{\text{pri},i}\left[k\right]$ is initialized by step 1 and updated by step 3. Then, the decoded LLR vector is obtained by
$\left({\bm \lambda}_{\text{dec}}\left[k\right],f_{\text{crc}}\right)={\sf ChannelDecoding}\left({\bm \theta}\left[k\right]\right)$,
the estimation ${\mathbf {\hat b}}\left[k\right]$ of ${\mathbf b}\left[k\right]$ is decided by ${\bm \lambda}_{\text{dec}}\left[k\right]$ and $\mathbf{\hat{q}}\left[k\right]$ is recovered from ${\mathbf {\hat b}}\left[k\right]$.
If $f_{\text{crc}}$ is $0$, the state vector $\mathbf{\tilde{x}}\left[k\right]$ and the input vector ${\mathbf u}\left[k\right]$ are calculated by Algorithm \ref{AlgorithmKF} and \eqref{EqInputVector}, respectively. If $f_{\text{crc}}$ is $1$, we use ${\bm \lambda}_{\text{dec}}\left[k\right]$ and KF to update ${\bm \theta}_{\text{pri}}\left[k\right]$ in step 3 to implement iterative detection.

When continuous decoding errors occur, the estimated probability density of ${\mathbf y}\left[k\right]$
is away from the real value, which further deteriorates the control and communication performance. To mitigate the error propagation, if the maximum iteration number $I$ is reached and $f_{\text{crc}}$ is $1$, a conventional channel decoding $\left({\bm \lambda}_{\text{dec}}\left[k\right],f_{\text{crc}}\right)={\sf ChannelDecoding}\left({\bm \theta}_{\text{rec}}\left[k\right]\right)$ is used, since ${\bm \theta}_{\text{rec}}\left[k\right]$ is related to the i.i.d. AWGN and received signals.

{\bf Step 3: Updating} ${\bm \theta}_{\text{pri}}\left[k\right]$. We have the probability
$
P\left(\left.b_j\left[k\right]\right|{\mathbf r}\left[k\right]\right) = \left(e^{\left(2b_j\left[k\right]-1\right)\lambda_{\text{dec},j}\left[k\right]} + 1\right)^{-1}
$.
Given the set ${\mathcal Z} = \left\{\left.{\hat z}_l\right|l=0,1,\cdots,2^n-1\right\}$ of the midpoint ${\hat z}_l$, the probability of
${\hat{\mathbf q}}\left[k\right] \in {\mathcal Z}^{N_y}$ given ${\mathbf r}\left[k\right]$ is
\begin{equation}\label{EqProqLambda}
\begin{aligned}
  & P\left( \left. {\hat{\mathbf q}}\left[k\right] \right| {\mathbf r}\left[k\right] \right)=\prod\limits_{{{i}_{1}}\in \left[\!\left[ {{N}_{y}} \right]\!\right]}{P\left( \left. {{\hat q}_{{{i}_{1}}}\left[k\right]} \right|{\mathbf r}\left[k\right] \right)} \\
 & =\prod\limits_{{{i}_{1}}\in \left[\!\left[ {{N}_{y}} \right]\!\right]}{\prod\limits_{{{i}_{2}}=n-1,\cdots ,0}{P\left( \left. b_{{{i}_{2}}}^{\beta \left( {{\hat q}_{{{i}_{1}}}\left[k\right]} \right)} \right|{\mathbf r}\left[k\right] \right)}} \\
 &= \prod\limits_{j \in \left[\!\left[ K\right]\!\right]}P\left(\left.b_j\left[k\right]\right|{\mathbf r}\left[k\right]\right). \\
\end{aligned}
\end{equation}

Given $\hat{\mathbf q}\left[k\right]$, the estimated system states $\mathbf{\hat x}\left[k\right]$ and the covariance matrix $\mathbf{\hat P}_x\left[k\right]$ are
\begin{equation}\label{EqIJDEstimatedXP}
\left(\mathbf{\hat x}\left[k\right], \mathbf{\hat P}_x\left[k\right]\right) = {\sf KF}\left(\mathbf{\tilde{x}}\left[ k-1 \right], \mathbf{\tilde{P}}\left[ k-1 \right], \hat{\mathbf q}\left[k\right], 0\right).
\end{equation}
The estimated sensor output is $
\mathbf{\hat y}\left[k\right] = \mathbf{C}\mathbf{\hat x}\left[k\right]$ and the corresponding covariance matrix is $
\mathbf{\hat P}_y\left[k\right] = \mathbf{C}\mathbf{\hat P}_x\left[k\right]\mathbf{C}^T + \mathbf{V}$.
Thus, $\mathbf{y}\left[k\right]$ given $\hat{\mathbf q}\left[k\right]$ follows the Gaussian distribution with mean $\mathbf{\hat y}\left[k\right]$ and covariance matrix $\mathbf{\hat P}_y\left[k\right]$ as
\begin{align}\label{EqProYgivenXk1Step3}
\left.\mathbf{y}\left[k\right]\right|\hat{\mathbf{q}}\left[k\right] \sim \mathcal{N}\left(\mathbf{\hat y}\left[k\right], \mathbf{\hat P}_y\left[k\right]\right).
\end{align}
The probability of quantized bit $b_{i_2}^{\beta \left( \hat q_{i_1}\left[k\right] \right)} \in \left\{0, 1\right\}$ is
\begin{equation}\label{EqProInfoBitStep3}
\begin{aligned}
  & P\left( b_{{{i}_{2}}}^{\beta \left( {{y}_{{{i}_{1}}}\left[k\right]} \right)} \right)
 =\sum\limits_{\hat{\mathbf{q}}\left[k\right]\in {\mathsf{\mathcal{Z}}}^{N_y}}{P\left( \hat{\mathbf{q}}\left[k\right]  \right)P\left( \left. b_{{{i}_{2}}}^{\beta \left( {{y}_{{{i}_{1}}}\left[k\right]} \right)} \right|\hat{\mathbf{q}}\left[k\right] \right)} \\
 & = \sum\limits_{{\hat{\mathbf{q}}\left[k\right]} \in {\mathcal{Z}^{{N_y}}}} P\left( {\hat{\mathbf{q}}\left[k\right]} \right) \times\\
 &\int\limits_{{y_{{i_1}}\left[k\right]} \in \mathbb{R}} {P\left( {\left. {b_{{i_2}}^{\beta \left( {{y_{{i_1}}\left[k\right]}} \right)}} \right|{y_{{i_1}}\left[k\right]}} \right)p\left( {\left. {{y_{{i_1}}\left[k\right]}} \right|{\hat{\mathbf{q}}\left[k\right]}} \right)} d{y_{{i_1}}\left[k\right]},
\end{aligned}
\end{equation}
where $P\left( \hat{\mathbf{q}}\left[k\right] \right)$ is set as $P\left( \left. \hat{\mathbf{q}}\left[k\right] \right| {\mathbf r}\left[k\right] \right)$ by \eqref{EqProqLambda}.
The calculation processes of $p\left( {\left. {{y_{{i_1}}}\left[k\right]} \right|{\hat{\mathbf{q}}\left[k\right]}} \right)$ and $P\left( {\left. {b_{{i_2}}^{\beta \left( {{y_{{i_1}}\left[k\right]}} \right)}} \right|{y_{{i_1}}\left[k\right]}} \right)$ are similar to \eqref{EqpYiXk1} and \eqref{EqpbjYi0}, respectively.
With \eqref{EqProInfoBitStep3}, we calculate ${{\lambda }_{\text{pri},j}}\left[k\right]$ by \eqref{EqPriorLLRbjInitial} and update ${\theta}_{\text{pri}, i}\left[k\right]$ by \eqref{EqPriorLLRciCal}
to implement iterative detection in step 2.
To reduce the complexity, we traverse $n_{\lambda}$ quantized bits with the least values of $\left|\lambda_{\text{dec},j}\right|$, $j\in\left[\!\left[ K
\right]\!\right]$, and use the $2^{n_{\lambda}}$ results to decide $P\left( \left. \hat{\mathbf{q}} \left[k\right] \right|\mathbf{r} \left[k\right] \right)$ with normalization and calculate \eqref{EqProInfoBitStep3}.

\begin{algorithm}[t]
\setlength{\abovecaptionskip}{0.cm}
\setlength{\belowcaptionskip}{-0.cm}
\caption{Iterative joint detection}\label{AlgorithmIJD}

\KwIn {The received LLR vector ${\bm \theta}_{\text{rec}}\left[k\right]$, the maximum iteration number $I$, the traversed bit number $n_{\lambda}$, the estimated system states $\mathbf{\tilde{x}}\left[ k-1 \right]$ and the covariance matrix $\mathbf{\tilde{P}}\left[ k-1 \right]$;}
\KwOut {The estimated state vector $\mathbf{\tilde{x}}\left[k\right]$ and the covariance matrix $\mathbf{\tilde{P}}\left[k\right]$;}

Initialize $i_{\text {ite}} = 1$\;

$\left(\mathbf{\hat x}\left[k\right], \mathbf{\hat P}_x\left[k\right]\right) = {\sf KF}\left(\mathbf{\tilde{x}}\left[ k-1 \right], \mathbf{\tilde{P}}\left[ k-1 \right], \mathbf{0}, 1\right)$\;

$\mathbf{\hat y}\left[k\right] = \mathbf{C}\mathbf{\hat x}\left[k\right]$  and $
\mathbf{\hat P}_y\left[k\right] = \mathbf{C}\mathbf{\hat P}_x\left[k\right]\mathbf{C}^T + \mathbf{V}$\;

Initialize $\theta_{{\text {pri}},i}\left[k\right]$ by \eqref{EqPriorLLRciCal}\;

\While{$i_{\textup{ite}} \le I$}
{
    $i_{\text {ite}} = i_{\text {ite}} + 1$ and ${{\bm\theta }}\left[k\right] = {{\bm\theta }_{\text{pri}}\left[k\right]}+{{\bm\theta }_{\text{rec}}\left[k\right]}$\;

    $\left({\bm \lambda}_{\text{dec}}\left[k\right],f_{\text{crc}}\right)={\sf ChannelDecoding}\left({\bm \theta}\left[k\right]\right)$\;

    \If{$f_{\textup{crc}}$ is $0$}
    {
        Recover $\mathbf{\hat q}\left[k\right]$ from ${\bm \lambda}_{\text{dec}}\left[k\right]$\;

        \Return $\left(\mathbf{\tilde{x}}\left[k\right], \mathbf{\tilde{P}}\left[k\right]\right) = {\sf KF}\left(\mathbf{\tilde{x}}\left[ k-1 \right], \mathbf{\tilde{P}}\left[ k-1 \right], \mathbf{\hat q}\left[k\right], 0\right)$\;
    }
    \Else
    {
        Initialize $T_{i_1,i_2} = 0$, ${i_1} \in \left[\!\left[ N_y
        \right]\!\right]$, ${i_2} = n-1,\cdots,0$\;

        Traverse $n_{\lambda}$ quantized bits with the least $\left|\lambda_{\text{dec},j}\right|$, $j\in\left[\!\left[ K
\right]\!\right]$ and the set of the $2^{n_{\lambda}}$ results is $\mathcal Q$\;

        \For{$\hat{\mathbf{q}}\left[k\right] \in \mathcal{Q}$}
        {
            $\left(\mathbf{\hat x}\left[k\right], \mathbf{\hat P}_x\left[k\right]\right) = {\sf KF}\left(\mathbf{\tilde{x}}\left[ k-1 \right], \mathbf{\tilde{P}}\left[ k-1 \right], \hat{\mathbf{q}}\left[k\right], 0\right)$\;

            $\mathbf{\hat y}\left[k\right] = \mathbf{C}\mathbf{\hat x}\left[k\right]$,
            $
            \mathbf{\hat P}_y\left[k\right] = \mathbf{C}\mathbf{\hat P}_x\left[k\right]\mathbf{C}^T + \mathbf{V}$\;

            Calculate $P\left( \left. \hat{\mathbf{q}} \left[k\right] \right|\mathbf{r} \left[k\right] \right)$ by \eqref{EqProqLambda}\;

            \For{${i_1} \in \left[\!\left[ N_y \right]\!\right]$, ${i_2} =n-1, \cdots, 0$}
            {
                $P_{i_1,i_2} = P\left( \left. b_{i_2}^{\beta \left( {y_{i_1}\left[k\right]} \right)} = 0 \right|\hat{\mathbf{q}}\left[k\right] \right)$\;

                ${{T}_{{{i}_{1}},{{i}_{2}}}}={{T}_{{{i}_{1}},{{i}_{2}}}}+\frac{P\left( \left. \widehat{\mathbf{q}}\left[ k \right] \right|\mathbf{r}\left[ k \right] \right)}{\sum\nolimits_{\widehat{\mathbf{q}}\left[ k \right]\in \mathsf{\mathcal{Q}}}{P\left( \left. \widehat{\mathbf{q}}\left[ k \right] \right|\mathbf{r}\left[ k \right] \right)}}{{P}_{{{i}_{1}},{{i}_{2}}}}$\;
            }
        }
        ${{\lambda }_{\text{pri},j}\left[k\right]}=\ln \frac{T_{i_1,i_2}}{1-{T_{{i_1},{{i}_{2}}}}}$ and set  $\theta_{{\text {pri}},i}\left[k\right]$ by \eqref{EqPriorLLRciCal}\;
    }
}

$\left({\bm \lambda}_{\text{dec}}\left[k\right],f_{\text{crc}}\right)={\sf ChannelDecoding}\left({\bm \theta}_{\text{rec}}\left[k\right]\right)$ and recover $\mathbf{\hat q}\left[k\right]$ from ${\bm \lambda}_{\text{dec}}\left[k\right]$\;

$\left(\mathbf{\tilde x}\left[k\right], \mathbf{\tilde P}_x\left[k\right]\right) = {\sf KF}\left(\mathbf{\tilde{x}}\left[ k-1 \right], \mathbf{\tilde{P}}\left[ k-1 \right], \mathbf{\hat q}\left[k\right], f_{\text{crc}}\right)$\;

\end{algorithm}

The whole process of the proposed iterative joint detection algorithm is summarized in Algorithm \ref{AlgorithmIJD}. In Algorithm \ref{AlgorithmIJD}, we first initialize ${{\bm\theta }_{\text{pri}}}\left[k\right]$ with \eqref{EqPriorLLRciCal} and calculate ${\bm \theta}\left[k\right]$ by \eqref{EqLLR}. Then, channel decoding is used as $\left({\bm \lambda}_{\text{dec}}\left[k\right],f_{\text{crc}}\right)={\sf ChannelDecoding}\left({\bm \theta}\left[k\right]\right)$. If $f_{\text{crc}}$ is $0$,  we assume the decoded results are correct and estimate $\mathbf{\tilde{x}}\left[k\right]$ and $\mathbf{\tilde{P}}\left[k\right]$ by KF directly. If $f_{\text{crc}}$ is $1$, we use ${\bm \lambda}_{\text{dec}}\left[k\right]$ and $\mathbf{\tilde{x}}\left[ k-1 \right]$ to update ${{\bm\theta }_{\text{pri}}}\left[k\right]$ to implement iterative detection until the maximum iteration number $I$ is reached. When $I$ is reached and $f_{\text{crc}}$ is $1$, we use a conventional channel decoding $\left({\bm \lambda}_{\text{dec}}\left[k\right],f_{\text{crc}}\right)={\sf ChannelDecoding}\left({\bm \theta}_{\text{rec}}\left[k\right]\right)$ to mitigate error propagation.

\section{Complexity and Time Step Analysis}

The complexity of the proposed iterative joint detection is shown in TABLE \ref{TableComplexity}. The complexity of step 1 is $O\left(N_x^3 + 2^nK\right)$, where the complexity of KF is $O\left(N_x^3\right)$ and the complexity of calculating the probability of quantized bits is $O\left(2^nK\right)$ when a lookup table is used for integrals.
Given the decoding complexity $O\left(D\right)$, the complexity of step 2 is $O\left(N_x^3 + D\right)$.
For step 3, the complexity is $O\left(2^{n_{\lambda}}\left(N_x^3 + 2^nK\right)\right)$.
Hence, the complexity of the iterative joint detection is $O\left(I\left(D + 2^{n_{\lambda}}\left(N_x^3 + 2^nK\right)\right)\right)$.
Then, $I = 1$ means only step 1 and step 2 are used and the complexity is $O\left(D + N_x^3 + 2^nK\right)$.
In comparison, the complexity of the detection in \cite{WNCSchannelCodingCRC} is $O\left(N_x^3 + 2^nK\right)$ and the difference lies in the additional decoding complexity $O\left(D\right)$.

Though the iterative joint detection has high complexity, it can be implemented in parallel for real-time systems.
The main time steps are shown in TABLE \ref{TableComplexity}.
Assume $\tau_{\sf kf}$ and $\tau_{\sf dec}$ are the time steps of KF and decoding, respectively. For step 1, since the time step of calculating the probability of one quantized bit is $n$ and $K$ bits can be calculated in parallel, the time step of step 1 is $\tau_{\sf kf}+n$. The time step of step 2 is $\tau_{\sf kf} + \tau_{\sf dec}$. Then, since the time step of \eqref{EqProqLambda} is $n_{\lambda}$ and the $2^{n_{\lambda}}$ results can be calculated in parallel, the time step of step 3 is $\tau_{\sf kf}+n+n_{\lambda}$. Thus, the time step of the iterative joint detection is $2\tau_{\sf kf}+n+\tau_{\sf dec}+I\left(\tau_{\sf kf} + \tau_{\sf dec} + n+n_{\lambda}\right)$
and a suitable $I$ should make the detection latency less than the sampling period.

\begin{table}[t]
\centering\caption{The complexity and the main time steps of the iterative joint detection}\label{TableComplexity}
\begin{tabular}{ccc}
\hline
Step                   & Complexity                                                                                            & Main Time Step                                                                   \\ \hline
1                      & $O\left(N_x^3 + 2^nK\right)$                                                                       & $\tau_{\sf kf}+n$                                                              \\
2                      & $O\left(N_x^3 + D\right)$                                                                            & $\tau_{\sf kf} + \tau_{\sf dec}$                                                  \\
3                      & $O\left(2^{n_{\lambda}}\left(N_x^3 + 2^nK\right)\right)$                                           & $\tau_{\sf kf}+n+n_{\lambda}$                                                              \\ \hline
\multirow{2}{*}{Total} & \multirow{2}{*}{$O\left(I\left(D + 2^{n_{\lambda}}\left(N_x^3 + 2^nK\right)\right)\right)$} & $2\tau_{\sf kf}+n+\tau_{\sf dec}+$                                                             \\
                       &                                                                                                       & \multicolumn{1}{l}{$I\left(\tau_{\sf kf} + \tau_{\sf dec} + n+n_{\lambda}\right)$} \\ \hline
\end{tabular}
\vspace{-1em}
\end{table}

\section{Simulation Results}

For the control layer, a rotary inverted pendulum with $Z = \pi$ is employed as the plant. The system parameter matrices $\mathbf{A}$, $\mathbf{B}$ and $\mathbf{C}$, the controller gain matrix $\mathbf{K}_\text{con}$ and the reference state vector ${{\mathbf{x}}_{{\text{ref}}}}\left[ k \right]$ are identical to those in \cite{WNCSchannelCodingCRC}.
$\mathbf{W}$ and $\mathbf{V}$ are diagonal matrices with the diagonal elements $\sigma_W^2$ and $\sigma_V^2$, respectively.
The sampling interval is $0.01$s, the simulation time is $100$s and the number of simulation runs is $100$.
For the communication layer, the outer code is a $\left(48, 32\right)$ 16-bit CRC code with $n = 16$ or a $\left(48, 24\right)$ 24-bit CRC code with $n = 12$ \cite{3GPP_5G_polar}. The inner code is a $\left(96, 48\right)$ LDPC code \cite{MacKayLDPC}. The traversed bit number $n_{\lambda}$ is 4.
The iteration number of BP decoding is 50. The comparison schemes are the CRC with MAP \cite{WNCSchannelCodingCRC} and the LDPC with BP \cite{MacKayLDPC}.

\begin{figure}[t]
\setlength{\abovecaptionskip}{0.cm}
\setlength{\belowcaptionskip}{-0.cm}
  \centering{\includegraphics[scale=0.63]{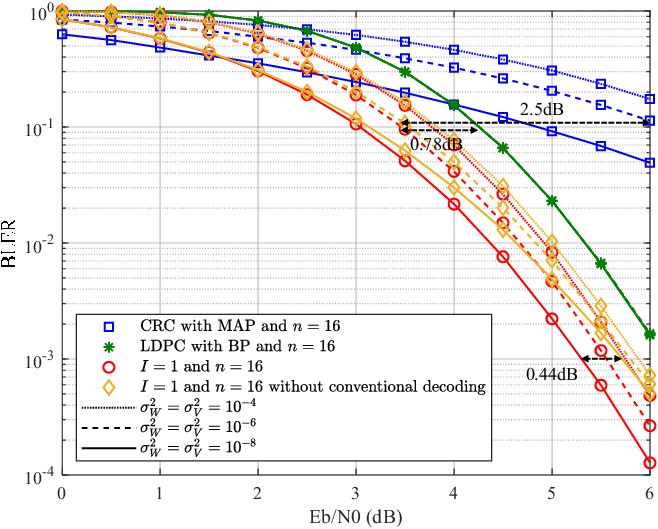}}
  \caption{The BLER performance of the iterative joint detection with $I = 1$ under different system disturbances.}\label{FigBLERI1}
  \vspace{-1em}
\end{figure}

Fig. \ref{FigBLERI1} provides the BLER performance of iterative joint detection with $I = 1$ under different system disturbances. 
In Fig. \ref{FigBLERI1}, for different $\sigma_W^2$ and $\sigma_V^2$, the proposed iterative joint detection outperforms the iterative joint detection without conventional decoding when $I$ is reached and the performance gap is about $0.44$dB at $\sigma_W^2 = \sigma_V^2 = 10^{-8}$ and BLER $10^{-3}$, 
which shows the advantages of utilizing the i.i.d. received signals to mitigate error propagation.
Then, as $\sigma_W^2$ and $\sigma_V^2$ decrease, the calculated prior probability becomes more accurate to improve the BLER performance of the iterative joint detection and \cite{WNCSchannelCodingCRC}.
Specifically, the iterative joint detection with $\sigma_W^2 = \sigma_V^2 = 10^{-6}$ has about $0.78$dB and $2.5$dB performance gain at BLER $10^{-1}$ compared with \cite{MacKayLDPC} and \cite{WNCSchannelCodingCRC}, respectively.
Hence, the prior information obtained from the system model can improve the BLER performance.

\begin{figure}[t]
\setlength{\abovecaptionskip}{0.cm}
\setlength{\belowcaptionskip}{-0.cm}
  \centering{\includegraphics[scale=0.63]{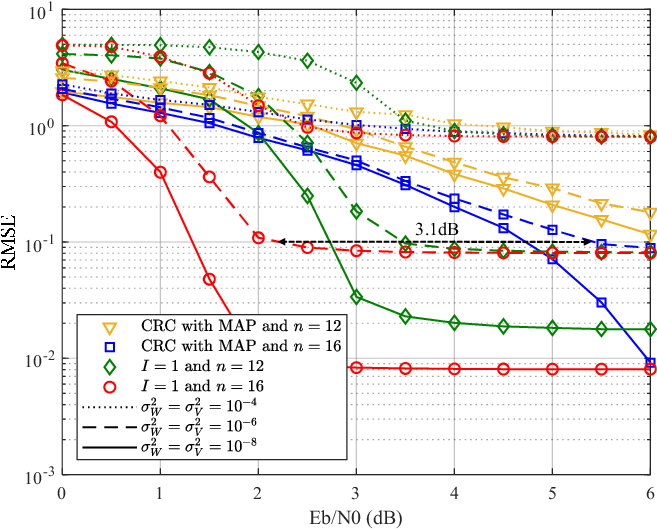}}
  \caption{The RMSE performance of the iterative joint detection with $I = 1$ under different system disturbances.}\label{FigRMSEI1}
  \vspace{-1.5em}
\end{figure}

Fig. \ref{FigRMSEI1} provides the RMSE performance of the iterative joint detection with $I = 1$ under different system disturbances.
In Fig. \ref{FigRMSEI1}, we observe that the RMSE under different system disturbances decreases and converges as the SNR increases. Then, the converged SNRs of the iterative joint detection are less than those of the CRC with MAP.
Specifically, the performance gap between the iterative joint detection with $n = 16$ and the CRC with MAP is about $3.1$dB with $\sigma_W^2 = \sigma_V^2 = 10^{-6}$ at RMSE $10^{-1}$.
Thus, the iterative joint detection can enhance the BLER performance of communications while improving the RMSE performance of controls, which shows the advantage of the joint design of controls and communications with prior information.
Additionally, since the quantization distortion with $n = 12$ is larger than that with $n = 16$, the converged RMSE with $n = 12$ under $\sigma_W^2 = \sigma_V^2 = 10^{-8}$ is larger. Thus, a higher quantization resolution can improve the RMSE performance under small system interference and selecting a suitable $n$ is a trade-off between performance and complexity.

\begin{figure}[t]
\setlength{\abovecaptionskip}{0.cm}
\setlength{\belowcaptionskip}{-0.cm}
  \centering{\includegraphics[scale=0.63]{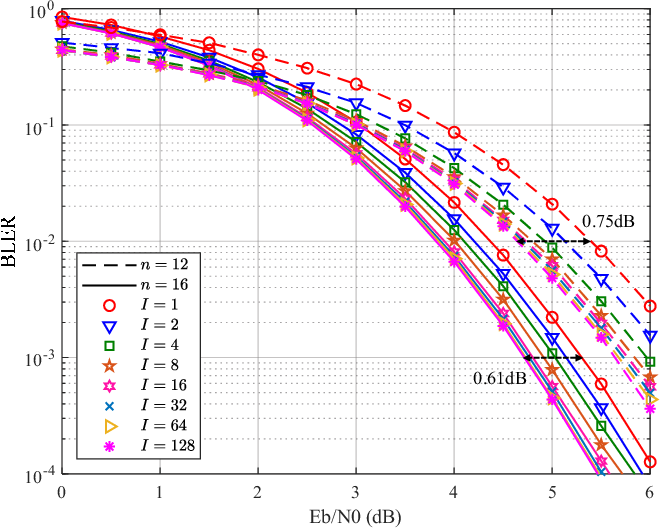}}
  \caption{The BLER performance of the iterative joint detection with different $I$ under $\sigma_W^2 = \sigma_V^2 = 10^{-8}$.}\label{FigBLER}
  \vspace{-1.5em}
\end{figure}

Fig. \ref{FigBLER} shows the BLER performance of the iterative joint detection with different $I$ under $\sigma_W^2 = 10^{-8}$, $\sigma_V^2 = 10^{-8}$.
In Fig. \ref{FigBLER}, we observe that as $I$ increases, the BLER performance is improved and converged.
As $I$ increases from 1 to 128, the performance gaps are about $0.61$dB and $0.75$dB at BLER $10^{-3}$ and $10^{-2}$ for $n = 16$ and $n = 12$, respectively.
Thus, updating the prior information can further improve the BLER performance of the iterative joint detection.

\section{Conclusion}

In this letter, an iterative joint detection algorithm of KF and channel decoder is proposed by utilizing the prior information of control systems to improve the control and communication performance.
In the algorithm, the prior information is initialized by KF and updated by traversing the possible outputs in order to implement iterative detection. The simulation results show that the prior information and the iterative structure can improve the control and communication performance.

\bibliographystyle{IEEEtran}
\bibliography{IEEEabrv,myrefs}

\end{document}